\documentclass[aps,twocolumn,showpacs]{revtex4}%
\usepackage{amsfonts}
\usepackage{amsmath}
\usepackage{amssymb}
\usepackage{graphicx}%
\begin{document}
\title{Experimental investigation of the origin of the cross-over temperature in the cuprates}
\author{Yuval Lubashevsky and Amit Keren}
\affiliation{Physics Department, Technion, Israel Institute of Technology, Haifa 32000, Israel}

\begin{abstract}
We investigate the cross-over temperature $T^{\ast}$ as a function of doping
in (Ca$_{x}$La$_{1-x}$)(Ba$_{1.75-x}$La$_{0.25+x}$)Cu$_{3}$O$_{y}$, where the
maximum $T_{c}$ ($T_{c}^{max}$) varies continuously by 30\% between families
($x$) with minimal structural changes. $T^{\ast}$ is determined by
DC-susceptibility measurements. We find that $T^{\ast}$ scales with the
maximum N\'{e}el temperature $T_{N}^{max}$ of each family. This result
strongly supports a magnetic origin of $T^{\ast}$, and indicates that three
dimensional interactions play a role in its magnitude.

\end{abstract}
\maketitle

Free electrons do not have high temperature cross-overs such as a pseudogap
(PG), spin gap (SG), or development of antiferromagnetic (AFM) correlations.
In the cuprates all of these exist, yet the interactions that lead to them are
not completely clear. Nevertheless, the cross-overs occur at a temperature
$T^{\ast}$ which is much higher than $T_{c}$, and closer to the three
dimensional (3D) ordering temperature of the parent compound in the AFM state.
Therefore, it is speculated that $T^{\ast}$ emerges from AFM fluctuations, and
that the cross-overs are intimately linked, namely, the interaction
responsible for one might be responsible for all
\cite{friend,Jperp,HarrisonPRL07}. Therefore, it is crucial to test the
possibility of correlations between $T^{\star}$ of a particular system and its
magnetic properties, such as the N\'{e}el temperature $T_{N}$ of the parent
compound, or its constituents, the in- and out-of-plane Heisenberg coupling
constant $J$ and $J_{\bot}$, respectively. This is the motivation of the work
presented here. We provide experimental evidence that strongly supports a
magnetic origin for $T^{\ast}$. Moreover, we show that $T^{\ast}$ stems from
3D interactions, similar to the N\'{e}el order, involving both $J$ and
$J_{\bot}$.

\begin{figure}
[h]
\begin{center}
\includegraphics[
width=\columnwidth
]%
{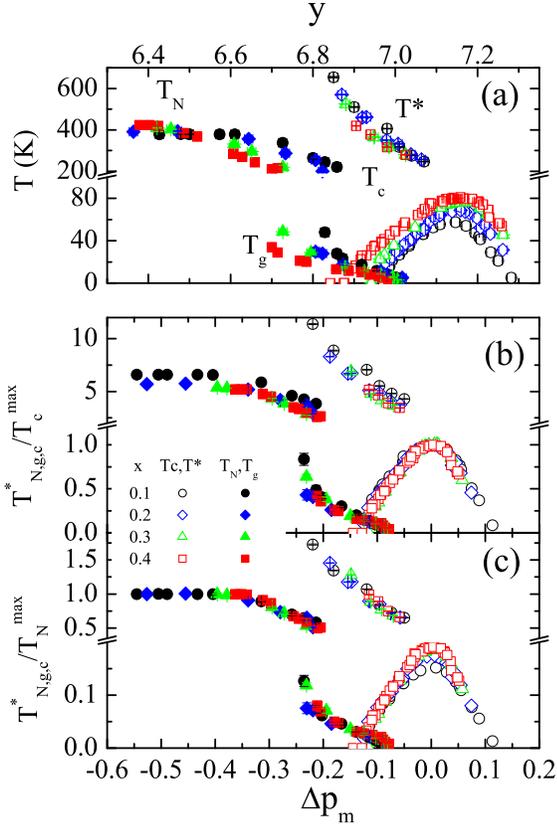}%
\caption{(Color online) \textbf{a}. The phase diagram of (Ca$_{x}$La$_{1-x}%
$)(Ba$_{1.75-x}$La$_{0.25+x}$)Cu$_{3}$O$_{y}$. \textbf{b}. The critical
temperatures are normalized by the maximum critical temperature $T_{c}^{max}$
of each family ($x$), and $y$ is replaced by mobile holes density variation
$\Delta p_{m}$ (see text). \textbf{c}. The same as b but the normalization is
by one number per family, referred to as $T_{N}^{max}$, which makes all
$T_{N}(x,y)$ curves collapse to one.}%
\label{CPD}%
\end{center}
\end{figure}

We investigate the origin of the $T^{\star}$ by studying its variations as a
function of the compound's magnetic properties, where small chemical changes
are an implicit parameter. The variations in the magnetic properties are
achieved by using four different families of the (Ca$_{x}$La$_{1-x}%
$)(Ba$_{1.75-x}$La$_{0.25+x}$)Cu$_{3}$O$_{y}$ (CLBLCO) system, having the
YBa$_{2}$Cu$_{3}$O$_{y}$ (YBCO) structure, with $x=0.1\ldots0.4$. The phase
diagram of the CLBLCO families is shown in Fig.~\ref{CPD}(a). $T_{c}$ was
measured by resistivity \cite{Goldschmidt}, and the spin glass temperature
$T_{g}$ \cite{KanigelPRL02} and $T_{N}$ \cite{Rinat} by muon spin
relaxation.\ Despite the rich phase diagram, the different CLBLCO families
have negligible structural differences. All compounds are tetragonal, and
there is no oxygen chain ordering as in YBCO \cite{Goldschmidt}. The hole
concentration in the CuO$_{2}$ planes does not depend on $x$
\cite{Chmaissem,KerenToBe}. The difference in the unit cell parameters $a$ and
$c/3$ between the two extreme families ($x=0.1$ and $0.4$) is
1\%~\cite{Goldschmidt}. Thus, variations in $T_{c}^{max}$ due to variations in
ionic radii are not relevant \cite{ChenPRB05}. The level of disorder, as
detected by Cu and Ca nuclear magnetic resonance, is also identical for the
different families \cite{KerenToBe,MarchandThesis}. In fact, the only strong
variation between families noticed at present is the in-plane oxygen buckling
angle \cite{OferToBe}. This property can modify the intraplane near- and
next-near-neighbors hopping, or interplane hopping, which controls the
magnetic interaction strengths $J$ and $J_{\bot}$ \cite{ttt}. The strong
magnetic and superconducting variations of the CLBLCO system, accompanied by
minimal structural changes, make it ideal for a correlation study between
$T^{\ast}$ and magnetism.%

\begin{figure}
[h]
\begin{center}
\includegraphics[
width=\columnwidth
]%
{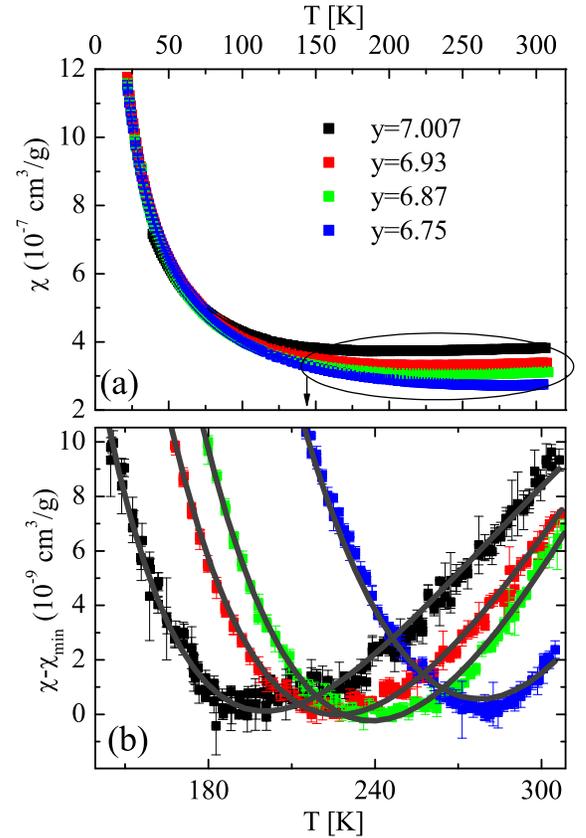}%
\caption{(Color online) \textbf{a}. Raw data from four samples of the $x=0.2$
family with a different doping levels. \textbf{b}. Zoom-in on the data in the
ellipse of panel a after the minimum value of $\chi$ is subtracted. The solid
lines are fits to Eq.~\ref{FitFunc}.}%
\label{data}%
\end{center}
\end{figure}

In this project we determine $T^{\ast}$ using temperature-dependent
magnetization measurements. In Fig.~\ref{data}(a) we present raw data from
four samples of the $x=0.2$ family with different doping levels. At first
glance the data contain only two features: A Curie-Weiss (CW) type increase of
$\chi$ at low temperatures, and a non-zero baseline at high temperature
$\chi_{300}$. This base line increases with increasing $y$. The CW term could
be a result of isolated spins, impurities, or spins on the chain layer.
However, as we will show shortly, there is much more to it. The baseline shift
could be a consequence of variations in the core and Van Vleck (CVV) electron
contribution or an increasing density of states at the Fermi level.

A zoom-in on the high temperature region, marked by the ellipse, reveals a
third feature in the data; a minimum point of $\chi$. To present this minimum
clearly we subtracted from the raw data the minimal value of the
susceptibility $\chi_{\min}$ for each sample, and plotted the result on a
tighter scale in Fig.~\ref{data}(b). The $\chi$ minimum is a result of
decreasing susceptibility upon cooling from room temperature, followed by an
increase in the susceptibility due to the CW term at low T. This phenomenon
was previously noticed by Johnston in YBCO \cite{Johnston0}, and Johnston and
Nakano \textit{et al.}~in La$_{2-x}$Sr$_{x}$CuO$_{4}$ (LSCO)
\cite{Johnson,Nakano}. The minimum point moves to higher temperatures with
decreasing oxygen level as expected from $T^{\ast}$. There are three possible
reasons for this decreasing susceptibility: (I) increasing AFM correlations
upon cooling \cite{Johnson}, (II) opening of a SG where excitations move from
$q=0$ to the AFM wave vector \cite{Neutrons}, or (III) disappearing density of
states at the Fermi level as parts of the Fermi arc are being gapped out when
the PG opens as $T/T^{\ast}$ decreases \cite{Kanigel}.

In order to determine the $T^{\star}$ we fit the data to a three component
function%
\begin{equation}
\chi=\frac{C_{1}}{T+\theta}+\frac{C_{2}}{cosh(\frac{T^{\star}}{T})}%
+C_{3}\text{.} \label{FitFunc}%
\end{equation}
The data are fitted without any restriction on the parameters. The quality of
the fit is demonstrated in Fig.~\ref{data}(b) by the solid lines; it captures
the data precisely with barely observable deviations at very low doping where
$T^{\ast}$ is at the edge of our measurement window. Of course on the broader
scale of Fig.~\ref{data}(a) there are no differences between the fit function
(not shown) and data. At dopings higher than $y\sim7.1$ the CW term overwhelms
the cross-over term due to its low $T^{\ast}$, and the $\chi$ minimum is no
longer detectable. At dopings lower than $y\sim6.85$ the $\chi$ minimum is out
of the measurement window. These samples are not analyzed. The $C$'s
determined by the fits (not shown) are found to behave smoothly and
monotonically as a function of doping and family. $C_{2}$ and $C_{3}$ have the
same order of magnitude as $\chi_{300}.$The function $\cosh^{-1}(T^{\star}/T)$
was chosen only because it fits the data best. However, we will analyze only
the scaling properties of $T^{\ast}$ for which the absolute value is not relevant.

As for the amplitudes, it is most natural to relate $C_{1}$ to the weight of
an impurity related CW contribution, $C_{2}$ to the crossed-over electrons,
and $C_{3}$ to free electrons and CVV susceptibility. This division is based
on the 2D Heisenberg model that predicts a decreasing susceptibility with
decreasing temperature~\cite{Assa}. However, the situation at hand is closer
to the t-J model for which the susceptibility is calculated by high
temperature series expansion, and its behavior at $T\rightarrow0$ is not known
\cite{SinghPRB92}. Therefore, it is conceivable that the division of $\chi$
into impurities, crossed-over, and free electron and CVV contribution is
artificial, that there is no impurities contribution, and that the
susceptibility simply has two energy scales $\theta$ and $T^{\star}$. We are
mostly interested in these two parameters and the evolution of $\chi_{300}$
with doping.%

\begin{figure}
[h]
\begin{center}
\includegraphics[
width=\columnwidth
]%
{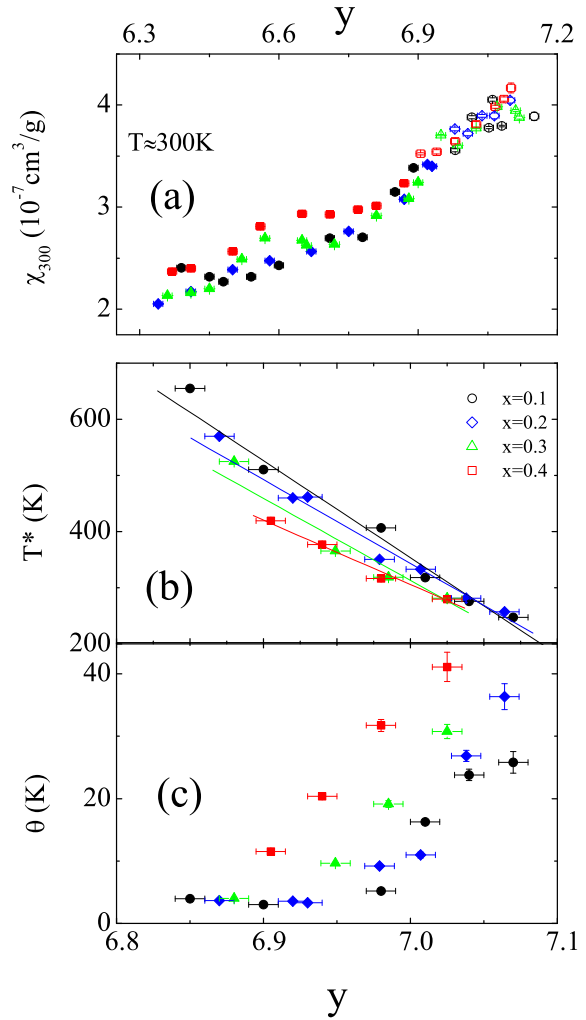}%
\caption{(Color online) \textbf{a}. The susceptibility of all the samples at
$T=300$~K. \textbf{b}. $T^{\star}$ as a function of doping and families. The
solid lines are guides to the eye. \textbf{c}. The Curie-Weiss temperature
$\theta$ as a function of doping and families.}%
\label{Y}%
\end{center}
\end{figure}

In Fig.~\ref{Y}(a) we plot $\chi_{300}$ for the different families. It is
clearly increasing as a function of $y$. The expected contribution from core
electrons, taken from the standard tables~\cite{selwood}, is also expected to
increase, but with a variation that is smaller in an order of magnitude. The
Van Vleck contribution is also taken as a constant \cite{Johnson}. Therefore,
the increasing of $\chi_{300}$ with doping must result from either an
increasing density of states at the Fermi level or decreasing correlation
length $\xi$. At very low doping, near the AFM phase, there are some
differences between the families; the $\chi_{300}$ is higher for the $x=0.4$
family. However, at doping level in which superconductivity appears,
$\chi_{300}$ is similar to all families. The density of states scenario is
consistent with previous claims that the doping level in CLBLCO is
$x$-independent, at least in the superconducting region
\cite{Goldschmidt,KerenToBe}. The correlation length scenario is not cosistent
with our previous claims that $J(x)$ varies by 30\% between families
\cite{Rinat} since $\xi$ has exponential $J$ dependence \cite{Assa}.

Since CLBLCO obeys the Uemura relation $T_{c}\propto n_{s}$ in the entire
doping range \cite{KerenSSC03}, where $n_{s}$ is the superconducting carrier
density, we conclude that the proportionality constant varies between
families, or that not all holes turn superconducting. This conclusion
reinforces our previous claims that $T_{c}\simeq J(x)n_{s}$ \cite{Rinat}, and
that in CLBLCO not all the holes condense to superfluid \cite{KerenToBe}.%

The $T^{\star}$ parameter obtained from the fits is depicted in Fig.~\ref{Y}%
(b) on a tight scale (the solid lines are a guide to the eye), and as part of
the full phase diagram in Fig.~\ref{CPD}(a). It behaves like the well-known PG
or SG $T^{\star}$ measured by other techniques on a variety of superconductors
samples \cite{friend}. At the same time, a decrease of $T^{\star}$ with doping
is consistent with the AFM correlation picture as a progressive departure from
the Mott insulator. More importantly, a small but clear family dependence of
$T^{\star}$ is seen. At first glance it appears that $T^{\star}$ has
anti-correlation with $T_{c}^{max}$ or the maximum $T_{N}$ ($T_{N}^{\max}$).
The $x=0.4$ family, which has the highest $T_{c}^{max}$ and $T_{N}^{\max}$,
has the lowest $T^{\star}$, and vice versa for the $x=0.1$ family.

However, this conclusion is reversed if instead of plotting the $T^{\star}$ as
a function of oxygen level, it is properly normalized, and plotted as a
function of mobile hole variation $\Delta p_{m}$. By mobile holes we mean
holes that \emph{do} turn superconducting as discussed above. $\Delta p_{m}$
is defined in two steps: I) The chemical doping measured from optimum, $\Delta
y=y-y_{0}$, is obtained for each compound ($y_{0}$ is the oxygen level at
$T_{c}^{\max}$). II) $\Delta y$ is multiplied by a different constant per
family $K(x)$, namely, $\Delta p_{m}=K(x)\Delta y$ \cite{Rinat}. The $K$'s are
chosen so that the superconductor domes, normalized by $T_{c}^{max}$ of each
family, collapse onto each other; $K=0.76,0.67,0.54,0.47$ with $5$\% accuracy
for $x=0.1\ldots0.4$ -see Fig.~\ref{CPD}(b)- .

\begin{figure}
[h]
\begin{center}
\includegraphics[
width=\columnwidth
]%
{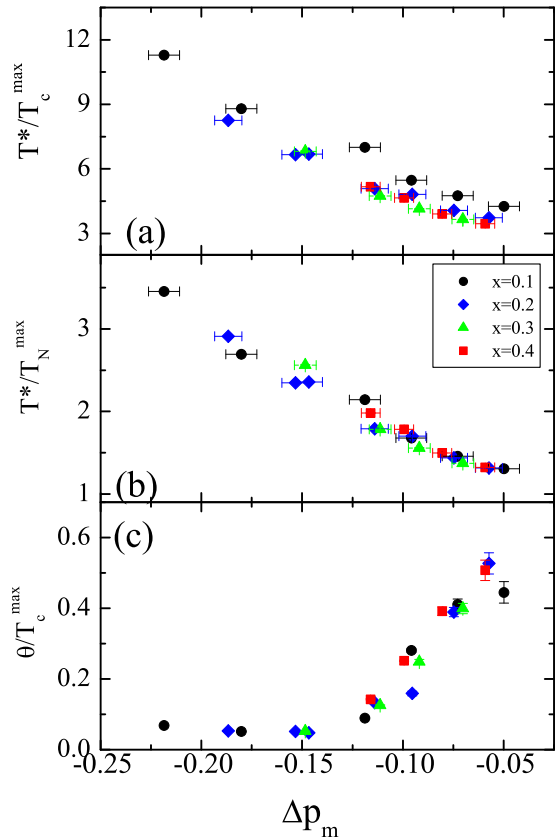}%
\caption{(Color online) \textbf{a}. $T^{\ast}/T_{c}^{max}$, \textbf{b}.
$T^{\ast}/T_{N}^{max}$, and \textbf{c. }the Curie-Weiss temperature\textbf{
}$\theta$, as a function of mobile hole variation $\Delta p_{m}$ (see text).}%
\label{DPm}%
\end{center}
\end{figure}

We examine two possible normalizations of the critical temperatures: by
$T_{c}^{max}$ or $T_{N}^{max}$. In Fig.~\ref{CPD}(b) we present \emph{all}
critical temperatures, normalized by $T_{c}^{max}$, as a function of $\Delta
p_{m}$. As expected, all domes scale onto each other. So do the glass
temperatures $T_{g}$. $T_{N}$ for $x=0.2,$ $0.3,$ and $0.4$ families also
collapse nicely. However, $T_{N}$ for the $x=0.1$ family does not. In
Fig.~\ref{DPm}(a)~we zoom in on the $T^{\ast}/T_{c}^{max}$, as a function of
$\Delta p_{m}$. The same problem is observed here as well. Next, we normalize
all critical temperatures by $T_{N}^{max}$ as shown in Fig.~\ref{CPD}(c). The
values of $T_{N}^{max}$ are chosen so that the $T_{N}(\Delta p_{m}%
)/T_{N}^{max}$ curves collapse onto each other, and are 379, 391.5, 410, and
423~K for the $x=0.1\ldots0.4$ families, respectively. Therefore, $T_{N}%
^{max}$ should be interpreted as the extrapolation of $T_{N}$ to the lowest
$\Delta p_{m}$ in Fig.~\ref{CPD}(c). In this case, the $T_{g}$ curves of all
family also collapse, but the $T_{c}$ domes do not. In Fig.~\ref{DPm}(b), we
zoom in on the $T^{\ast}$/$T_{N}^{\max}$, as a function of $\Delta p_{m}$. Now
all the normalized $T^{\ast}$ curves overlap. Thus $T^{\ast}$ of each family
scales better with $T_{N}^{\max}$ than with $T_{c}^{max}$. This is our main finding.%

As for the CW parameter $\theta$, although we did not expect any correlations
between this parameter and $x$ or $y$, we found an interesting trend shown in
Fig.~\ref{Y}(c). In the antiferromagnetic region $\theta\sim0$. As we go to
higher doping levels this magnetic energy scale increases. This trend was
previously observed by Bobroff \textit{et al.} \cite{BobroffPTL99}. It is also
clear that there are systematic variations of $\theta$ between the families.
The $x=0.4$ has the strongest $\theta$, and the $x=0.1$ the weakest. In fact,
in Fig.~\ref{DPm}(c) we plot $\theta/T_{c}^{max}$ as a function of $\Delta
p_{m}$. All data point collapse to a single curve. Once again we find that the
proper doping parameter is $\Delta p_{m}$, and not oxygen level $y$, and the
same energy scales control both $\theta$ and $T_{c}^{max}$. These trends
suggest that $\theta$ has nothing to do with impurities, as already hinted above.

Our results bare important new information on $T^{\star}$. When we normalize
$T^{\star}$ by $T_{c}^{max}$ we are actually normalizing by the in-plane
energy scale of each family $J(x)$ \cite{Rinat}. If the cross-over was only a
result of magnetic interaction between the spins in the planes (2D),
$T^{\star}$ should have scaled with $T_{c}^{max}$. The imperfect normalization
by $T_{c}^{max}$, demonstrated in Fig.~\ref{DPm}(a), contradicts this
possibility. When we normalize $T^{\star}$ by $T_{N}^{max}$, we are taking
into the account the coupling between the planes $J_{\bot}$ (3D) as well. The
success of the normalization by $T_{N}^{\max}$, shown in Fig.~\ref{DPm}(b),
implies that $T^{\star}$ is governed by 3D magnetic interaction.

The importance of $J_{\bot}$ was previously discussed in Ref.~\cite{Jperp}.
Our finding is also consistent with the concept of a 3D to 2D crossover above
$T_{c}$ in which planes decouple from each other \cite{PlaneDecoup}. Finally,
it is consistent with Nakano \textit{et al.} where by comparing LSCO to
Bi$_{2}$Sr$_{2}$CaCu$_{2}$O$_{8+\delta}$ a proportionality between $T^{\ast}$
and an unspecified magnetic energy scale is found \cite{NakanoJPSJ98}.

To summarize, after converting oxygen level $y$ to mobile holes variations
$\Delta p_{m}$, we find that the cross-over temperature $T^{\star}$ measured
by susceptibility in the CLBLCO system is proportional to $T_{N}^{max}$.
$T_{N}$ is set by both in- and out-of-plane coupling constants that are
determined by in- and out-of-plane hoppings. This result suggests a 3D
magnetic origin for $T^{\star}$. In addition, the CW-like term of the
susceptibility is not a result of impurities. It might be an intrinsic
property of doped CuO$_{2}$ planes at low temperatures.

We would like to acknowledge financial support from the Israel Science
Foundation and the Posnansky research fund in high temperature superconductivity.

\end{document}